\documentclass[12pt,preprint]{aastex}

\shorttitle{The photometric study of non-relaxed cluster "Mayer 3"}
\shortauthors{Tadross, A. L.$^a$, Bendary, R.$^a$, Priya, H.$^b$}

\begin{document}

\title{The photometric study of non-relaxed cluster "Mayer 3"}

\author{Tadross, A. L.$^a$, Bendary, R.$^a$, Priya, H.$^b$}
\affil{$^a$National Research Institute of Astronomy and Geophysics, 11421 - Helwan, Cairo, Egypt.\\
$^b$ Muffakham Jah College of Engineering and Technology, Hyderabad 500 034, India.}
\email{altadross@gmail.com}

\begin{abstract}
The astrophysical parameters of the open star cluster Mayer 3 have been estimated using the Newtonian focus (f/4.84) of the 1.88 m Telescope of Kottamia Observatory in Egypt. The VRI observations have been carried out down to a limiting magnitude of V$\sim$ 20 mag. To emphasize the main astrophysical parameters, Near-Infra Red, Two Micron All Sky Survey (2MASS-JHK) database has been used. Analyzing the color-magnitude diagrams and radial density distribution of Mayer 3, the cluster's age, reddening, distance from the Sun, the limited border, core and tidal radii are evaluated. The photometrical membership, the Galactocentric coordinates from the Galactic center and Galactic plane are estimated. Luminosity - mass functions, and the total mass of the cluster are established. Finally, the relaxation time estimation conclude that Mayer 3 is not dynamically relaxed yet.
\end{abstract}


\section{Introduction}

One of the most important components in studying the evolution and the structure of the Milky Way Galaxy is the open star clusters. More than two thousand such objects have been detected and catalogued. Although many of these objects are newly discovered and need photometric investigations, some others need the confirmation of their physical nature whether they are real star clusters or just asterism stars. Mayer 3 is a poorly studied cluster; its J2000.0 coordinates are $\alpha=7^{h} \ 30^{m} \ 6^{s}, \ \delta=-18^{\circ} \ 32^{'} \ 00^{''}, \ \ell= 233.7547^{\circ} \& \ b= -0.1868^{\circ}$. The first multicolor CCD photometric study of this object was published in 2013 used Himalayan Chandra Telescope, Hanle, India combined with the 2MASS JHK values, which has no photometric data published so far, Sujatha et al. (2013). In the present work, we present a real photometric study for Mayer 3 in optical VRI bands observed by the 1.88 m Telescope of Kottamia Astronomical Observatory (KAO) of Egypt. For more details about KAO, see Azzam et al. (2010). In addition, to confirm the photometrical main parameters, NIR photometry of PPMXL catalog of R\"{o}ser et al. (2010) has been used. It combines the USNO-B1.0 database of Monet et al. (2003) and the 2MASS database of Skrutskie et al. (2006) yielding the largest collection of proper motions in the International Celestial Reference Frame (ICRF). USNO-B1.0 contains the positions of more than one billion objects taken photographically around 1960, which gives us an opportunity to distinguish between the members and background field stars.\\

 The 2MASS survey provides J (1.25µm), H (1.65µm) and Ks (2.17µm) band photometry of millions of galaxies and nearly a half-billion stars
(Carpenter 2001).\\

Our aim in the present continuation series of papers is to determine the main astrophysical properties of poorly or/ and unstudied star clusters (cf. Tadross 2008a, 2008b \& Tadross 2009a, 2009b and what comes next). However, the essential astrophysical properties (distance, age, reddening, diameter, … etc.) besides, the structural parameters (limited cluster radius, core radius, tidal radius, Galactocentric coordinates and the distance from the Galactic plane) have been estimated here for the first time. The present study concludes that Mayer 3 is not relaxed yet.

 This paper is arranged as follows: Observation and data reduction are given in section 2. Radial density profile is described in Sections 3. Color-Magnitudes diagrams analysis are declared in Section 4. Luminosity, mass functions and the dynamical status of the cluster are estimated in Section 5. Finally, the conclusion of the present study is given in section 6.

\section{Observations and Data Reduction}

The CCD-VRI photometric observations of Mayer 3 have been carried out with the 1.88 m Reflector Telescope of the KAO, Egypt. The observation has been secured in the Newtonian focus with a plate scale of 22.53 arcsec/mm and field area of 10$\times$10 arcmin$^2$ on the night 7-8 July 2014. Fig. (1) represents the {\it VRI-image} of the cluster obtained from the 1.88 m Reflector of KAO of Egypt (left panel); a {\it JHK-image} taken from the 2MASS database (right panel), which shows a nebulosity and a marked concentration in the cluster. The characteristics of the CCD Camera used in KAO are explained in Table (1).

\begin{table*}
\centering
\caption{The characteristics of CCD Camera used in the observations.}
 \begin{tabular}{cl}
  \hline\noalign{\smallskip}
Type & EEV CCD 42-40                    \\
Version & Back-illuminated with BPBC (Basic Processes Broadband Coating)  \\ 
Format  & 2048 $\times$ 2048 pixel$^2$                  \\
Pixel size   & 13.5 $\times$ 13.5 $\mu$m$^2$                 \\
Grade  & 0                  \\
Dynamic range  & 30: 1                 \\
A/D converter  & 16 bit                  \\
Imaging area  & 27.6 $\times$ 27.6 mm$^2$               \\
Read out noise @20 KHz & 3.9 e$^-$ / pixel                \\
Gain   & 2.26 e$^-$ /ADU (Left amplifier) \& 2.24 e$^-$ /ADU (Right amplifier)                 \\
  \noalign{\smallskip}\hline
\end{tabular}
\end{table*}

For calibration purposes, the standard sequence Landolt (1992) PG 1323-086 has been observed in the VRI filters during the same night from which the extinction coefficients can be obtained. Table (2) summarizes the different VRI observations of Mayer 3 and the standard sequence of PG 1323-086.

The observed CCD frames are treated for bias and flat field corrections using the standard procedures of {\it IRAF} software and the photometry carried out using {\it DAOPHOT} package (Stetson 1987, 1992). The standard field star PG 1323-086 (Landolt, 1992) is observed for standardization. Extinction coefficients and zero points are determined to standardize the data. The magnitude errors in each band is presented in Fig. (2). The numbers of observed stars in V, R, and I bands are 547, 921, 809, respectively (with errors $<$ 0.1 mag).

To obtain the instrumental magnitudes of the stars in the observed filed, the Point-spread function (PSF) of Stetson (1987) has been used. For calibration process, the standard field PG 1323-086 of Londolt (1992) is observed in the same nights. The frames have been treated and corrected using aperture photometry to obtain the instrumental magnitude of the standard stars. The extinction coefficients and the zero point for each color have been calculated to standardize the instrumental magnitudes of the cluster stars applying the transformation equations:

\begin{center}
v = V + Z$_v$ + K$_v$ X + a$_v$ (B-V)\\
r = R + Z$_r$ + K$_r$ X + a$_r$ (V-R)\\
i = I + Z$_i$ + K$_i$ X + a$_i$ (V-I)\\
\end{center}

Where v, r, and i are the instrumental magnitudes; V, R, and I are the standard magnitudes; while Z, K, and a are the zero points, the extinction coefficients, and the color coefficients in VRI bands respectively, which their values are listed in Table (3); X is the air mass. In low galactic latitudes, the clusters are not sufficiently prominent more than the background field stars, so it is difficult to defined the cluster members easily. Because we didn't have a comparison field in our VRI observations, the photometric criterion for excluding the field stars is applied. A star regards as a member if it lies inside the cluster limited area and, in the same time, falls within the photometric envelope (close to the main sequence curve), i.e. about $\pm$ 0.15 mag along the abscissa of color-axis, (Clari$\acute{a}$ \& Lapasset 1986).
\\

To reduce the background contamination to some extent, the field density distribution is achieved for an offset field sample as following. The offset field sample here is an external ring, which has the same area of the studied cluster lies in Northern declination, one degree away from the cluster's center. Avoiding spatial variations in the number of faint numerous stars, which affected by large errors (Bonatto et al. 2004), the cutoff J $<$ 16.0 mag is applied a to both cluster and offset field stars as well.

The proper motions and JHKs photometric data have been extracted from the published catalog of the PPMXL Survey of R\"{o}ser et al. (2010), which included in an area of 10 arcmin centered on the cluster. The raw data of the cluster has been given using the VizieR\footnote{\it http://vizier.cfa.harvard.edu/viz-bin/VizieR?-source=I/317} tool. Applying the completeness limit (J $<$ 16.5 mag) to the cluster stars (Bonnato, et al., 2004), for the photometric quality, the stars with magnitude errors $\geq$ 0.2 mag have been excluded.
\\

Sampedro et al. (2017) used positions and proper motions of the fourth United States Naval Observatory (USNO) CCD Astrograph Catalog (UCAC4) with 3-different methods to determine membership probabilities for each star in 1876 clusters. For Mayer 3, they identified 9 members and found the mean propoer motions $\mu{RA}=0.59 \pm 0.88$ mas/yr and $\mu{\delta}=1.1 \pm 0.40$ mas/yr. Proper motion vector point diagram (VPD) with distribution histogram of 5 mas/yr bins for (pm $\alpha$ cos $\delta$) and (pm $\delta$) have been constructed as shown in Fig. 3. The Gaussian function fit to the central bins provides the mean pm in both directions, which are found in good agreement with Sampedro et al. (2017) and to some extent with GAIA-survey DR1. All data that lie at mean $\pm 1~\sigma$ (where $\sigma$ is the standard deviation) can be considered as probable members. All of which are marked on the CMDs.
\\

The Two Micron All Sky Survey is uniformly scanning the entire sky in three near-IR bands J (1.25 $\mu$m), H (1.65 $\mu$m) and Ks (2.17 $\mu$m) with two highly automated 1.3 m telescopes equipped with a three channel camera, each one consists of a 256 $\times$ 256 array of HgCdTe detectors. The photometric uncertainty of the data is less than 0.155 mag with a photometric completeness of Ks $\sim$ 16.5 mag. At a given magnitude, the errors are affected for Ks band (Soares \& Bica 2002). Then, the raw data of J and H bands with small errors are used to probe the fainter stars of Mayer 3.

\begin{table*}
\centering
\caption{The observation log of Mayer 3 and Landolt PG 1323-086.}
  \begin{tabular}{clcl}
  \hline\noalign{\smallskip} 			
Object & Filter & Exp. Time (Sec.) & No. of frames \\
  \hline\noalign{\smallskip}
         & V & 360 & 9 \\
Mayer 3  & R & 240 & 9 \\
         & I & 30 & 5 \\
         & I & 60 & 3  \\
         & I & 240 & 9  \\
 \hline\noalign{\smallskip}
            & V & 60& 8\\
Landolt PG 1323-086 & R & 100 & 4 \\
            & I & 100 & 4 \\
  \noalign{\smallskip}\hline
\end{tabular}
\end{table*}

\begin{table}
\centering
\caption{The obtained values of the zero point, the extinction and color coefficients.}
\begin{tabular}{clcl}
  \hline\noalign{\smallskip}
Filter &  Z  & K & a  \\
  \hline\noalign{\smallskip}
V & 2.755 & 0.152 & 0.001   \\
R & 2.643 & 0.087 & 0.001   \\
I & 3.382 & 0.043 & -0.003   \\
  \noalign{\smallskip}\hline
\end{tabular}
\end{table}

\section{Radial Density Profile}

The stars inside Mayer 3 area, which are taken from the PPMXL database, counted in concentric rings to outward from the cluster center with equal increment radius of 0.1 arcmin. Fig. 4 represents the radial density distribution of the cluster, where the mean stellar density in each ring is plotted against the corresponding mean radius. The cluster border is taken at that limit which contains the whole cluster area and reaches a sufficient stability with the background field density, where the cluster stars dissolved in the background field ones (cf. Tadross, 2005). Applying the empirical King's Model (1966), the density function $\rho(r)$ can be represented as:

\begin{equation}
\rho(r)=f_{bg}+\frac{f_{0}}{1+(r/r_{c})^{2}}
\end{equation}

where $f_{bg}$, $f_{0}$ and $r_{c}$ are the background field, central densities and the core radius of the cluster respectively.
\\
They are found to be 13.7 star/arcmin$^2$, 0.30 star/arcmin$^2$ and 0.46 arcmin respectively. From Fig. (4) the angular limited radius is found to be 4.7 arcmin, which corresponds to distance of 3.3 pc.

\section{Colour-Magnitude Diagrams analysis}

The Color-Magnitude Diagrams (CMDs) for the observed stars in the cluster region V$\sim$(V-I), V$\sim$(V-R), and R$\sim$(R-I) are plotted as shown in Fig. (5). The blue dotes refer to the stars lie within the cluster limited boundaries. These observed CMDs have been compared to many theoretical Padova isochrones (Girardi et al. 2010) with different ages of solar mitallicity Z = 0.019. Applying the normal fitting method, the best fit is obtained at age of 4 Myr and distance modulus of 12.0$\pm$0.10 mag. The reddening values are found to be 0.84, 0.20 and 0.38 mag respectively. The ratio AV/E(B-V) is taken to be 3.1 (Garcia et al. 1988).

On the 2MASS CMDs, J$\sim$(J-H) and K$\sim$(J-K) are plotted for Mayer 3 as shown in Fig. (6). The blue dotes refer to the stars lie within the cluster boundaries, while the red ones represent the offset field background sample. Using theoretical isochrones of Girardi et al. (2010) of the same age; the best fit is obtained at distance modulus (m-M) = 12.0 $\pm$ 0.10 mag. The color excesses E(J-H) \& E(J-K) are found to be 0.27 and 0.44 mag respectively. These reddening values are corresponding to an optical color excess of E(B-V) = 0.88 mag.
\\
For absorption transformation between the JHK and VRI systems, we used the ratio AJ/AV = 0.278 and AH/AV = 0.176 (Schlegel et al., 1988) and AK/AV = 0.118 (Dutra et al., 2002). According to the photometric observations and the 2MASS data, the distance of Mayer 3 is found to be 2510$\pm$110 pc. Correspondingly, the mean linear diameter, the distance from galactic center, from the galactic plane, the distances X and Y from the Sun on the galactic plane are found to be 6.6 pc, 10.2 Kpc, -8.0 pc, 1478 pc, and -2016 pc respectively, (cf. Tadross, 2011).

\section{Luminosity, Mass Functions and the dynamical status}

The luminosity and mass functions (LF \& MF) depend on the membership of the cluster. Using the photometric data and the positions of the cluster stars, the stars close to the central region of the cluster and lie on/and close to the main sequence curves of the CMDs, within a photometric uncertainty of $\pm$ 0.15 mag, considered as probable members. Accordingly, a number of 805 stars in the cluster region have been counted inside the limit diameter (6.6 pc). The apparent magnitude of each star has been transformed to absolute magnitude using the obtained distance modules. The absolute magnitude has been classified to bins of 0.5 mag for each. The counted number of stars in each bin has been used in constructing the luminosity function as shown in Fig. (7). The magnitude bin intervals are selected to include a reasonable number of stars in each bin and for the best possible statistics of the LF and MF. In this context, the total luminosity of the cluster is found to be $\sim$ --7.5 mag.

The LF and the MF are correlated to each other according to the known mass-luminosity relation. Converting the magnitudes to masses using the theoretical model of Girardi et al. (2010), the mass function has been constructed as shown in Fig. (8). The initial mass function (IMF) can be represented as:

\begin{equation}
\frac{dN}{dM} \propto M^{-\alpha}
\end{equation}

where $\frac{dN}{dM}$ is the number of stars in the mass interval [M:(M+dM)] and $\alpha$ is the slope of the mass function. Applying the linear fit to the mass function distribution, the slope of the relation is derived as -2.42$\pm$0.45, which is in a good agreement with Salpeter (1955). We calculated the total mass of the cluster by integrating the masses of the probable members (Sharma et al. 2006), which is found to be 1260 $M_{\odot}$.\\
\\
Knowing the total mass of the cluster, applying the equation of Jeffries et al. (2001), the tidal radius can be given as:

\begin{equation}
R_{t} = 1.46 (M_{c})^{1/3} ,
\end{equation}
where $R_{t}$ and $M_{c}$ are the tidal radius and total mass of the cluster respectively. $R_{t}$ is calculated to be 15.8 pc.
\\
\\
Following Spitzer \& Hart (1971), the dynamical relaxation time (T$_R$) could be obtained from the relation:
\begin{equation}
\large T_{R} = \frac{8.9 \times 10^{5} \sqrt{N} \times R_{h}^{1.5}}{\sqrt{m} \times log (0.4 N)}
\end{equation}

where $N$ is the number of the cluster members, $R_{h}$ is the radius containing half of the cluster mass in parsecs, and $m$ is the average mass of the cluster in solar unit, assuming that $R_{h}$ equals half of the cluster radius. Applying the previous equation, the relaxation time is found to be 18 Myr (more than the cluster's age). This mean that Mayer 3 is not dynamically relaxed yet. Maybe the reason for that situation due to the imprint of star formation or the dynamical evolution of the very young system or both.

\section{Conclusions}

The poorly studied open star cluster Mayer 3 has been observed using the optical CCD-VRI bass-band of the 1.88 m Kottamia Telescope of Egypt. All the astrophysical parameters of this cluster have been estimated and confirmed photometrically by the 2MASS-JHK database analysis. It is noted that the calculation of the cluster limited diameter using PPMXL database allow us to obtain reliable data for large extension to the clusters' halos. From the CMDs it is clear that this cluster has a clear main sequence branch of age 4 Myr. The best fitting of the cluster with the isochrones are obtained at the same distance modulus of 12.0$\pm$0.10 mag, which corresponds to distance of 2510$\pm$110 pc and different reddening, which are in agreement and correlated to each other. Also, the mass function distribution verified that the IMF slope of Mayer 3 agrees with Salpeter (1955)' value within the error. The relaxation time of the cluster compared to its age indicate that Mayer 3 is not dynamically relaxed yet.

\begin{figure} \resizebox{\hsize}{!}
{\includegraphics[]{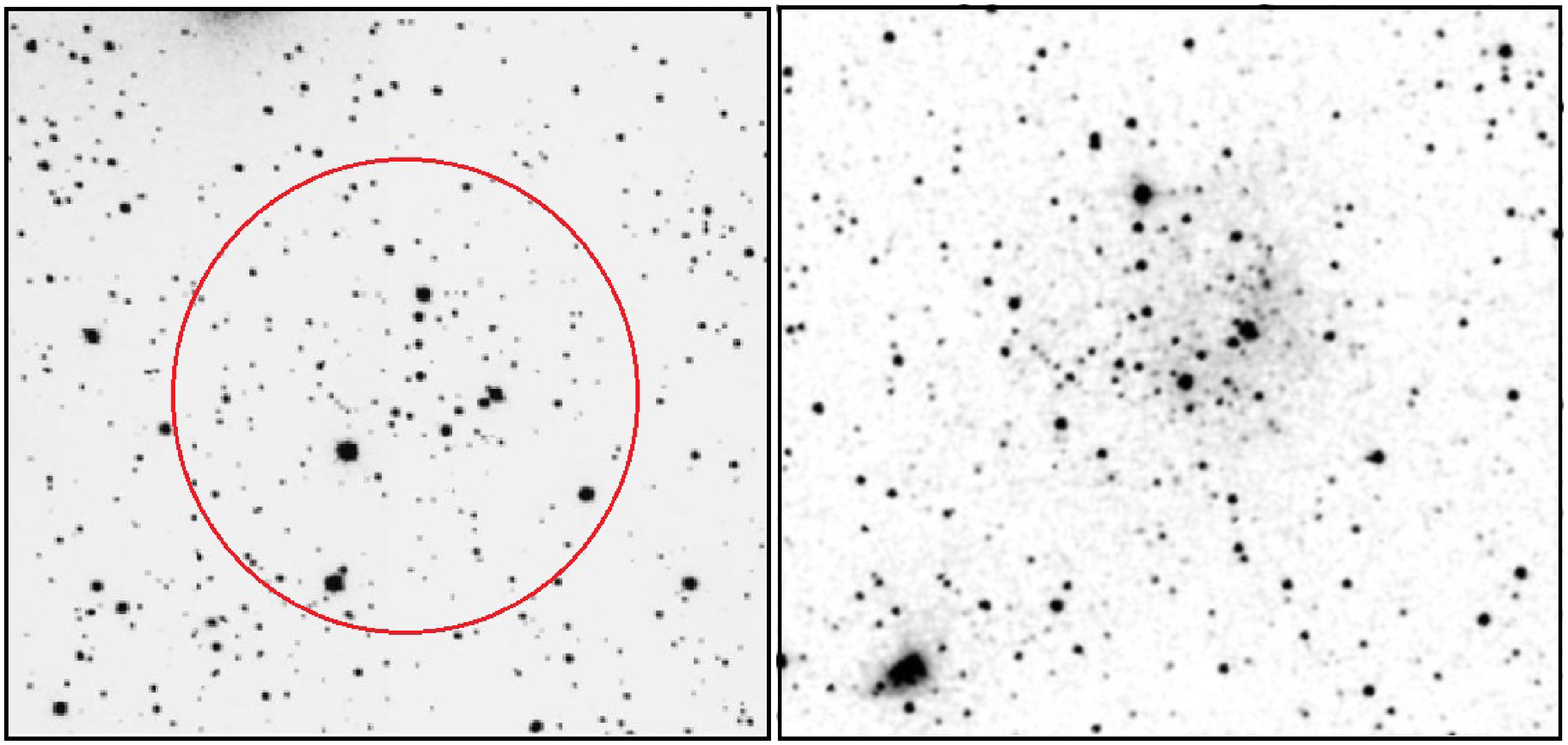}}
\caption{Left Panel: the {\it VRI-image} of Mayer 3 as observed by 1.88 m Reflector of KAO of Egypt, the red circle refers to the roughly diameter of the cluster. Right Panel: the {\it JHK-image} as taken from 2MASS database, a nebulosity and a marked concentration can be shown. North is up, East on the left.}
\label{label1}
\end{figure}

\begin{figure}\resizebox{\hsize}{!}
{\includegraphics[]{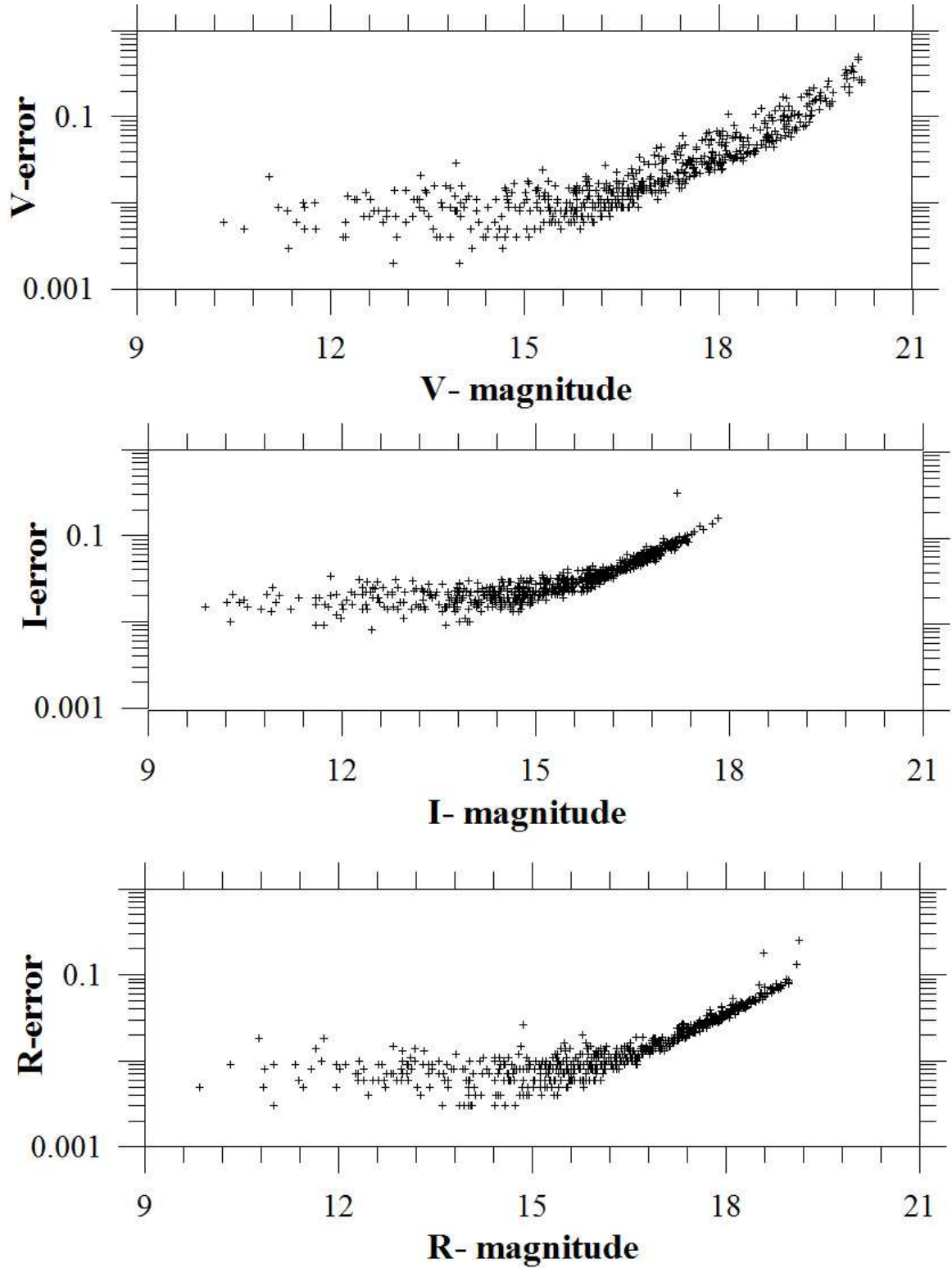}}
\caption{{\it VRI} errors of the observed magnitudes for the stars of Mayer 3.}
\label{label1}
\end{figure}

\begin{figure}\resizebox{\hsize}{!}
{\includegraphics[]{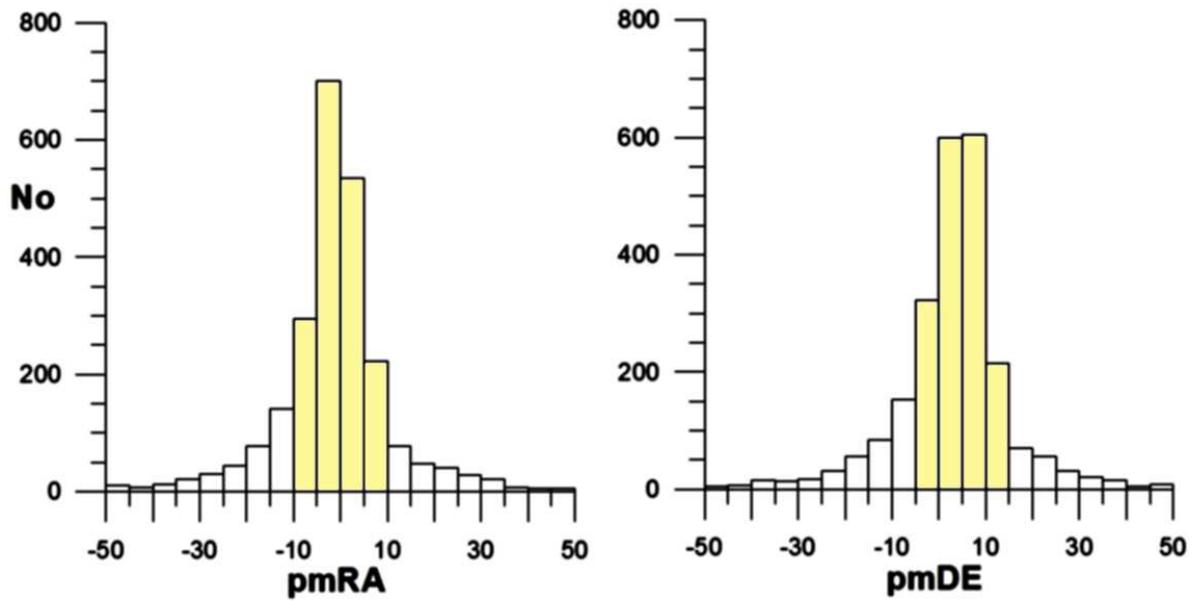}}
\caption{Proper motion histograms of 5 mas/yr bins in right ascension and declination of Mayer 3. The shaded regions represent the limits at which the membership taken. The Gaussian function fit to the central bins provides the mean values in both directions, which are found in good agreement with Sampedro et al. (2017). The standard error is taken to be $\pm 1\sigma$.}
\end{figure}

\begin{figure}\resizebox{\hsize}{!}
{\includegraphics[]{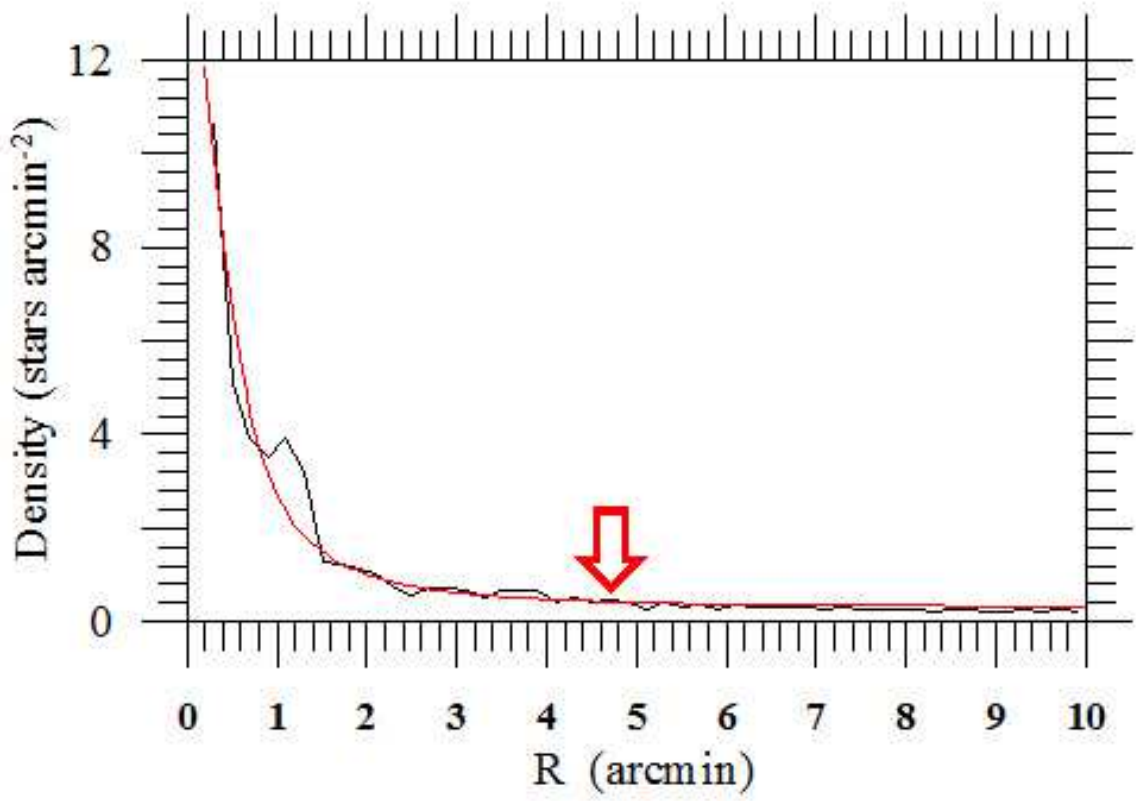}}
\caption{The radial density profile of Mayer 3. The model curve of King (1966) has been applied.}
\label{label1}
\end{figure}

\begin{figure}\resizebox{\hsize}{!}
{\includegraphics[]{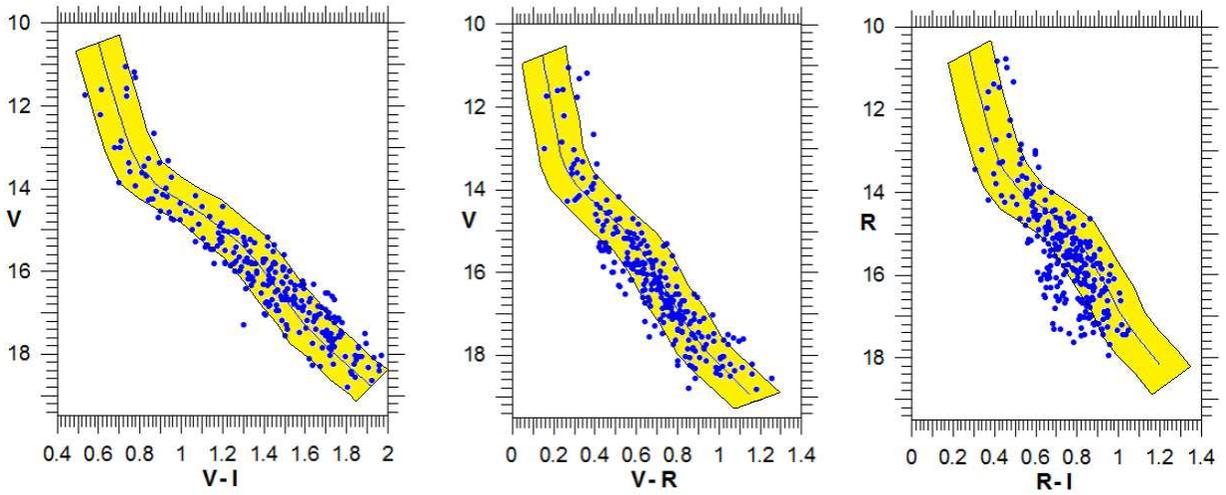}}
\caption{The observed CMDs of Mayer 3 fitted to the theoretical isochrones of Girardi et al. (2010). The distance modulus is found to be 12.0 mag, and the reddening values are found to be (from left to right) 0.84, 0.20 and 0.38 mag respectively. The blue dotes refer to the stars lie within the cluster edges. The yellow areas represent the photometric envelopes along the main sequence curves, i.e. $\pm$ 0.15 mag around the color-axis.}
\label{label1}
\end{figure}

\begin{figure}\resizebox{\hsize}{!}
{\includegraphics[]{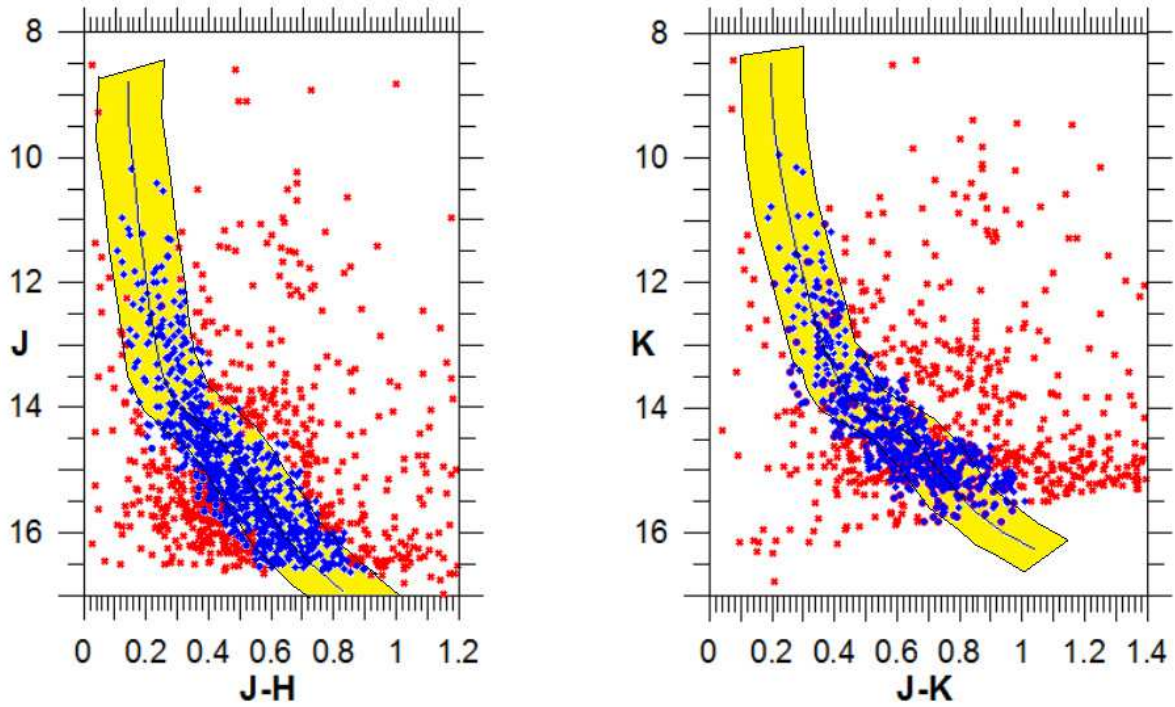}}
\caption{The 2MASS CMDs of Mayer 3 fitted to the theoretical isochrones of Girardi et al. (2010). The distance modulus is found to be 12.0 mag, and the color excesses are found to be (from left to right) 0.27, 0.44 mag respectively. The blue dotes refer to the stars lie within the cluster edges, while the red ones represent the stars outside. The yellow areas represent the photometric envelopes along the main sequence curves, i.e. $\pm$ 0.15 mag around the color-axis.}
\label{label1}
\end{figure}

\begin{figure}\resizebox{\hsize}{!}
{\includegraphics[]{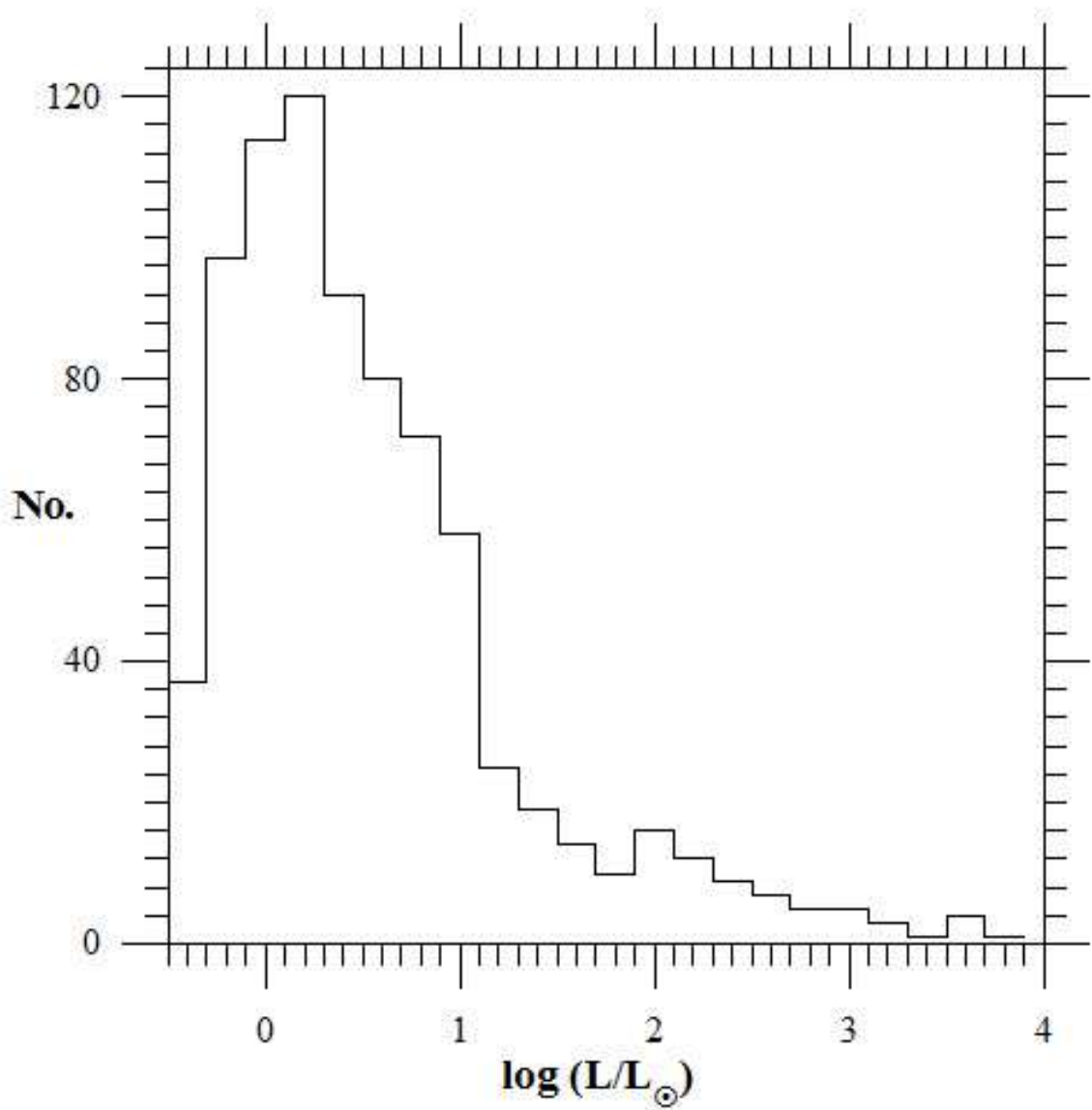}}
\caption{The luminosity function of Mayer 3.}
\label{label1}
\end{figure}

\begin{figure}\resizebox{\hsize}{!}
{\includegraphics[]{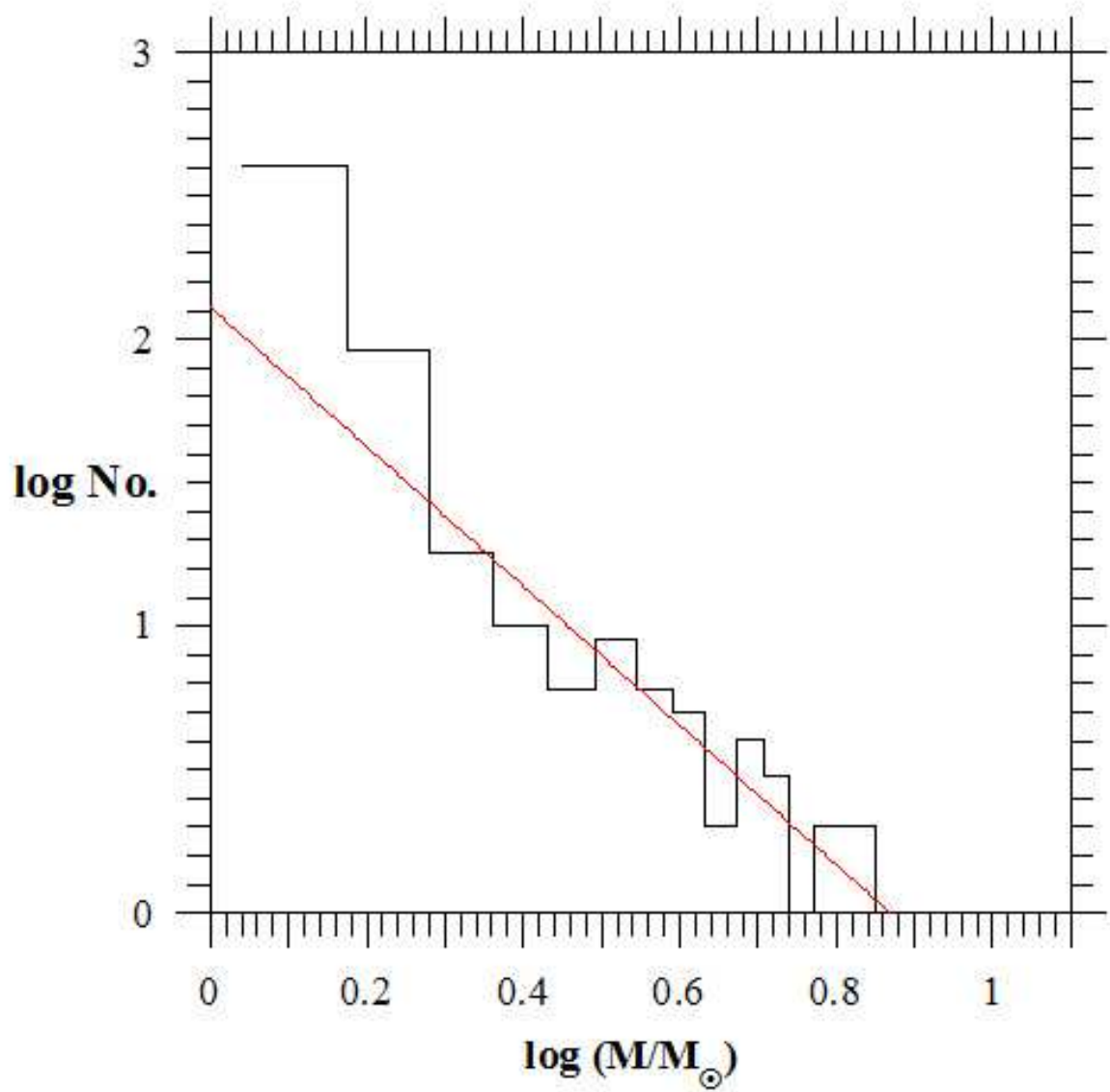}}
\caption{The mass function of Mayer 3.}
\label{label1}
\end{figure}
\end{document}